\documentclass[%
  aps,prl,
  superscriptaddress,
  amsmath,amssymb,
  reprint
]{revtex4-1}

\usepackage{epstopdf}
\usepackage{graphicx}
\usepackage{dcolumn}
\usepackage{bm}
\usepackage{siunitx}

\begin{document}
\bibliographystyle{apsrev4-1}

\title{Observing zero-field spin dynamics with spin noise in a pi-pulse modulated field} 
\author{Guiying Zhang}
\affiliation{Institute of Modern Physics, Department of Nuclear Science and Technology and Applied Ion Beam Physics Laboratory, Key Laboratory of the Ministry of Education, Fudan University, Shanghai 200433, China}
\affiliation{College of Science, Zhejiang University of Technology, Hangzhou 310023, China}
\author {Ya Wen}
\affiliation{Institute of Modern Physics, Department of Nuclear Science and Technology and Applied Ion Beam Physics Laboratory, Key Laboratory of the Ministry of Education, Fudan University, Shanghai 200433, China}
\author {Jian Qiu}
\affiliation{Insitute for Electric Light Sources, School of information Science and Engineering, Fudan University, Shanghai 200433, China}
\author {Kaifeng Zhao}
\email[]{zhaokf@fudan.edu.cn}
\affiliation{Institute of Modern Physics, Department of Nuclear Science and Technology and Applied Ion Beam Physics Laboratory, Key Laboratory of the Ministry of Education, Fudan University, Shanghai 200433, China}

\date{\today}
\begin{abstract}
Spin noise spectroscopic study of spin dynamics in a zero magnetic field is commonly masked by the dominating $1/f$ noise. We show that in a pi-pulse modulated magnetic field, spin noise spectrum centered at one-half of the modulation frequency reveals spin dynamics in a zero-field free of any low-frequency noise.
\end{abstract}
\maketitle 
A sample of $N$ paramagnetic spins at thermal equilibrium generates spin fluctuations of the order of $\sqrt{N}$. According to the fluctuation-dissipation theorem \cite{kubo_fluctuation-dissipation_1966}, the power spectrum of fluctuations, whether they are classical or quantum in nature, is proportional to the frequency response of the system to a small driving force, and vice versa. The measurement of such fluctuations, spin noise spectroscopy, was pioneered in the early 1980s \cite{aleksandrov_magnetic-resonance_1981,sleator_nuclear-spin_1985}. With the modern instrumental advancement such as real-time spectrum analyzer and ultra-fast digitizer, it has become a powerful non-perturbative way to obtain information about the spin dynamics of various systems including atomic vapors \cite{crooker_spectroscopy_2004}, semiconductors \cite{oestreich_spin_2005,muller_semiconductor_2010,hubner_rise_2014}, and quantum dots \cite{crooker_spin_2010,li_intrinsic_2012}.  The most efficient non-perturbative spin noise detection method is by measuring the Faraday rotation (FR) of an off-resonant linearly polarized beam passing through a strictly unpolarized sample \cite{zapasskii_spin-noise_2013,sinitsyn_theory_2016}. FR is also widely used for calibrating the spin noise for quantum metrology, where an unpolarized system is more favorable than a polarized one which is prone to converting classical noises into spin fluctuations or to introduce back-action noises, both of which scale as $N$ and overwhelm the projection noise \cite{sorensen_quantum_1998,vasilakis_stroboscopic_2011}.

In conventional FR spin noise measurements, a DC magnetic field transverse to the probing direction is applied to shift spin noise spectra (SNS) to a high-frequency region free from any technical noises, especially the dominating $1/f$ noise \cite{aleksandrov_magnetic-resonance_1981}. But spin dynamics in (near) zero-field can be very different from that in large fields \cite{happer_spin-exchange_1973,dellis_spin-noise_2014,dahbashi_optical_2014,berski_interplay_2015}. Cross-correlation SNS has been used to study heterogenous interacting spin system in zero-field. \cite{roy_cross-correlation_2015}.  Although SNS in zero-field can be obtained by subtracting SNS in zero and high transverse field from each other \cite{dahbashi_optical_2014}, due to the wandering $1/f$ noise, such approach only works when the linewidth or the power of the spin noise is much larger those of the $1/f$ noise. Here we show that by using a $\pi$ pulse modulated (PM) field, one can observe the spin dynamics in zero-field with SNS signals shifted to one-half of the field modulation frequency. We demonstrate this technique by studying the spin-exchange relaxation (SER) in a $^{87}$Rb atomic vapor, which consists of two spin species with equal but opposite g-factors. We also propose a generalized $\pi$-PM field for studying interactions between spins with an arbitrary ratio of g-factors. Suppressing SER by pulsed modulated field was first demonstrated with $2\pi$-PM \cite{korver_suppression_2013}, which has no time-averaged spin precession. $\pi$-PM with synchronous $\sigma+$ and $\sigma-$ pumping has been used to achieve SERF magnetometry and to suppress the back-polarization field of 87Rb in a Helium-Neon comagnetometer \cite{limes_3he-129xe_2018}. It should be noted that a very similar technique was independently developed by Zhang and Zhao to achieve SERF Bell-Bloom magnetometry in large fields with phase sensitive detection \cite{zhang_atomic_2016} following their proposal of using $\pi$-PM field to evade light-shift-back-action-noise \cite{zhang_quantum_2016}. SNS in sinusoidal bias fields has been studied theoretically \cite{braun_faraday-rotation_2007} on the 2nd-order spin correlations and experimentally on the 4th-order spin correlations \cite{li_higher-order_2016}. Interestingly, ultra-high frequency SNS can be shifted to the low-frequency region with a pulsed probe laser \cite{muller_gigahertz_2010,starosielec_ultrafast_2008}.
\begin{figure}
  \includegraphics[width=8.6cm]{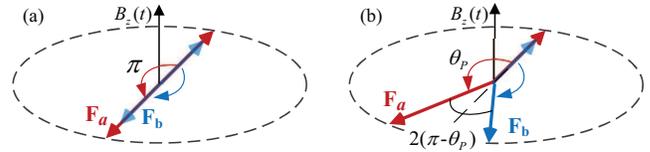}
  \caption{\label{fig:ArbitraryPulseArea}(color online) Spin precessions on the two hyperfine sublets in a $\pi$-PM (a) or an arbitrary $\theta_{p}$-PM field (b).}
\end{figure}

The basic idea of our method is shown in Fig.\ref{fig:ArbitraryPulseArea}(a). The ground state of alkali metal atoms has two hyperfine sublets with corresponding atomic spin numbers $F_a=I+1/2$ and $F_b=I-1/2$, where $I$ is the nuclear spin number. The two sublets have equal but opposite g-factors, thus $\mathbf{F}_a$ and $\mathbf{F}_b$ precess in the opposite direction under a constant magnetic field. Without loss of generality, assuming both spins are in the same direction initially, their relative angle, as well as the magnitude of the total spin changes coherently with time. If a random spin exchange collision (SEC) takes place between a pair of atoms during the precession, while the total spin is conserved, each atom flips between the two hyperfine sublets, reversing its spin precession direction. Thus their coherence is destroyed and the magnitude of the total spin relaxes. If we replace the constant field by a $\pi$-PM one and assume that the pulse-width is so narrow that SECs can only take place between the pulses, then each pulse rotates $\mathbf{F}_a$ by an angel of $\pi$ and $\mathbf{F}_b$ by $-\pi$, making them pointing in the same direction at the end. The internal state of the atomic system is the same before and after the pulse, except for an overall spatial rotation which has no effect on the microscopic SECs. Therefore, such a PM field is equivalent to a zero-field for SEC even though each spin reverses its direction pulse after pulse.

We first give a simple theory for the SNS in a PM field. According to the Wiener-Khintchine theorem, the power spectral density (PSD) of a random spin fluctuation, $F(t)$, is equal to the Fourier transform of the $F(t)$'s autocorrelation function, $C(\tau) \equiv \langle F(t)F(t+\tau)\rangle$,
\begin{equation}\label{eq:WienerKhintchine}
  S(\nu)= 2 \int_{0}^{\infty}\cos(2\pi\nu \tau) C(\tau) d\tau.
\end{equation}
For spins with a constant relaxation rate in a bias magnetic field, $C(\tau)$ is given by
\begin{equation}\label{eq:CTau}
  C(\tau) = \langle F_0^2 \rangle e^{-\Gamma |\tau|} \cos [ \theta(\tau) ],
\end{equation}
where $\Gamma$ is the transverse relaxation rate, $\theta(\tau)$ is the spin precession angle during time $\tau$ and $\langle F_0^2 \rangle=F(F+1)/3$ is the variance of spin noise. Under a DC field of strength $B$, $\theta(\tau)=2\pi\nu_{L} \tau$, where $2\pi\nu_{L}=g\mu_{B}B$ is the Larmor frequency with  $\mu_B$ being the Bohr magneton. Assuming $\Gamma<<2\pi\nu_L$, we have
\begin{equation}\label{eq:PSD}
   S(\nu)= \frac{\langle F_0^2 \rangle}{2\pi} \frac{\delta/2}{(\nu-\nu_L)^2 +\delta^2/4}.
\end{equation}
where $\delta = \Gamma/\pi$ is the full-width-half-maximum (FWHM). If the bias field is modulated with sharp pulses at frequency $\nu_p$, $\theta(\tau)$ can be approximated by a staircase function,
  $\theta(\tau) = \lfloor (\tau+\tau')/T_{p} \rfloor \theta_{p}$,
where $T_{p} = 1/\nu_{p}$ is the time interval between successive pulses, $\theta_{p}$ is the pulse area which is the spin precession angle caused by each pulse, $\tau'$ is the time delay between the occurrence of a random spin fluctuation and the last pulse, and $\lfloor x \rfloor$ is the floor function which gives the largest integer $\leq x$. Since spin fluctuation is a stationary random process, $\tau'$ should be distributed between $(0,T_{p})$ uniformly \cite{braun_faraday-rotation_2007}. Thus
\begin{align}\label{eq:CTauPulsedField}
  C(\tau) =  \frac{1}{T_{p}}\int_{0}^{T_{p}} d\tau' \langle F_0^2 \rangle e^{-\Gamma |\tau|}\cos[\lfloor (\tau+\tau')/T_{p} \rfloor \theta_{p}].
\end{align}
When $\Gamma \ll \bar{\nu}_{L}$, where $2\pi\bar{\nu}_{L} = \theta_{p}/T_{p}$ is the average Larmor frequency of the pulsed field, we find
\begin{equation}\label{eq:PSDArbitraryPulse}
   S(\nu) \propto \sum_{n=1,3,5}^{\infty}\sum_{s=-}^{+} \frac{1-\cos(\theta_{p})}{T_{p}^2 \nu_{n,s}^2}\frac{\delta/2}{(\nu-\nu_{n,s})^2+\delta^2/4},
\end{equation}
where $\nu_{n,\pm} = [n \pm (\theta_{p}-\pi)/\pi] \nu_{p}/2$ represent the center frequency of each resonance. Note, $n$ sums over all the positive odd numbers. When $\theta_{p} \neq \pi $, each index $n$ corresponds to a doublet centered at $n\nu_{p}/2$ with a frequency splitting of $\nu_{n,+}-\nu_{n,-}=\nu_{p}(\theta_{p}-\pi)/\pi$ .
When $\theta_{p} = \pi $, the doublet merges into a single peak at $n \nu_{p}/2$, and
\begin{equation}\label{eq:PSDPiPulse}
   S(\nu) \propto \sum_{n=1,3,5}^{\infty} \frac{8}{n^2 }\frac{\delta/2}{(\nu-n\nu_{p}/2)^2+\delta^2/4}.
\end{equation}

\begin{figure}
  \includegraphics[width=8.6cm]{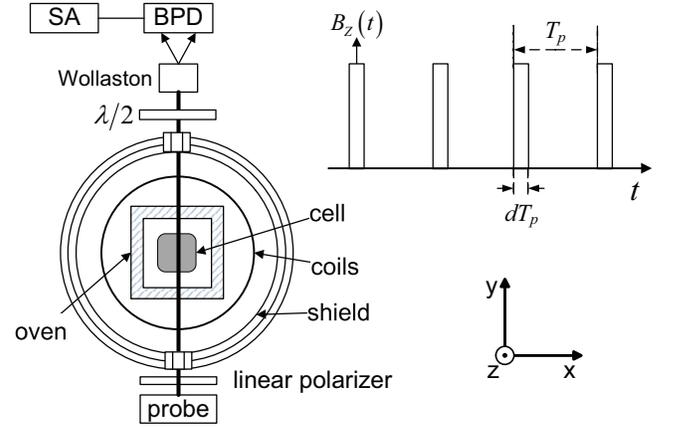}%
  \caption{\label{fig:Setup}(color online) Experiment setup and parameters of the PM field. BPD: balanced photo detector, Wollaston: Wollaston prism, $\lambda/2$: half-wave plate, $d$: duty cycle.}
\end{figure}

Our experimental set up is shown in Fig.\ref{fig:Setup}. A \SI{2.4}{cm} cubic Pyrex cell containing $^{87}$Rb with no buffer gas is placed in a ceramic oven. The inner wall of the cell is coated with octadecyltrichlorosilane \cite{zhang_effects_2015} to preserve the spin coherence over several hundred wall collisions.  The oven is heated by a nonmagnetic wire and is well insulated so that after the heating current is turned off, the temperature of the cell drops less than \SI{0.2}{K} within \SI{80}{s} of signal averaging time. A \SI{30}{cm} diameter Helmholtz coil controlled by a low noise current source (ADC6156) provides the constant field. The pulsed-field is provided by a \SI{18}{cm} diameter coil system driven by a homemade pulse current supply controlled by a function generator. The inductance of pulse coil is \SI{16}{\micro H}. The whole setup is enclosed by a four-layer $\mu$-metal shield to reduce the residue field below \SI{1}{nT} after degaussing. A \SI{1}{MHz} linewidth external cavity diode laser (Toptica DLpro) red-detuned \SI{1.8}{GHz} from 87Rb D1 line $F=2$ to $F'=1$ transition is used as the probe beam. The detuning is much larger than the Doppler broadening (\SI{300}{MHz}) and homogeneous broadening ($\sim\SI{10}{MHz}$) of the optical resonance, ensuring negligible photon absorption. However, it is much smaller than the ground state hyperfine splitting so that the FR signal is mostly contributed by the spin orientation from the atoms on the $F=2$ hyperfine sublet. The probe beam passes through a single mode optical fiber and is collimated into a beam of \SI{4}{mm} diameter. It is then linearly polarized before entering the vapor cell with a power of \SI{0.5}{mW}. The FR of the probe is measured with a polarizing beam splitter and a balanced photodetector (Thorlabs PDB210A) whose output is fed into a spectrum analyzer (SRS760) for PSD measurement. All spectra are linearly rms averaged $5000$ times. The time of averaging is about \SI{80}{s} for a \SI{1.6}{kHz} span and decreases with increasing span range.
\begin{figure}
  \includegraphics[width=8.6cm]{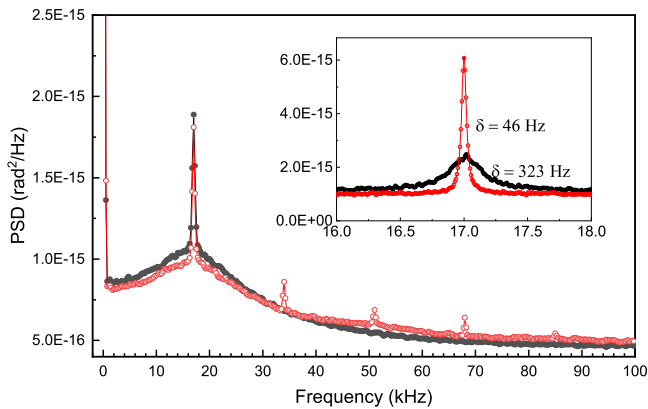}
  \caption{\label{fig:PSDofDCandPulsedField}(color online) SNSs under a DC field ($\nu_L=\SI{17}{kHz}$) (filled black dots) and a $\pi$-PM field ($\nu_p=\SI{34}{kHz}, d=2.5\%$) (empty red circles). The cell temperature is \SI{106}{\celsius}. The resolution of the spectrum analyzer for the main plot and the inset is \SI{250}{Hz} and \SI{8}{Hz}, respectively. For the DC field, the SNS peaks at \SI{17}{kHz} with a wide transit pedestal and a narrow Ramsey peak. For the PM field, the two peaks at 34 and \SI{68}{kHz}, corresponding to integer multiples of $\nu_p$, are parasitic signals stimulated by the magnetic pulses. The three peaks at 17, 51 and \SI{85}{kHz}, corresponding to half-integer multiples of $\nu_p$, are the SNS predicted by Eq.(\ref{eq:PSDPiPulse}). }
\end{figure}

Fig.\ref{fig:PSDofDCandPulsedField} compares the measured SNS in a DC and a $\pi$-PM field. The main plot shows the whole spectrum of a \SI{100}{kHz} span while The inset shows a zoomed scan around the first Larmor resonance. Under the DC field, the SNS centers at the Larmor frequency ${\nu}_{L}$ of the field, and contains a wide pedestal due to the transit time effect and a narrow peak due to the wall induced Ramsey effect. For the transit pedestal, the observed spin autocorrelation decays quickly as the atoms fly out of the probe region, while for the Ramsey peak, the observed spin autocorrelation lasts over a much longer time as atoms transverse the probe region many times due to coherence preserving collisions with the anti-relaxation cell wall. The linewidth of the Ramsey peak is mostly contributed by the SER broadening which is $q\sigma nv/\pi\sim $\SI{290}{Hz}, where $\sigma$ is the Rb SE cross-section $\sim $\SI{2e-14}{\cm\squared}, $n$ the Rb number density $\sim $\SI{0.85e12}{\per\cubic\cm}, $v$ the relative atomic speed $\sim $\SI{430}{m\per\s} and $q=1/8$ the nuclear slowing-down-factor for the $F=2$ sublevel \cite{happer_effect_1977}. The rest of about \SI{33}{Hz} is contributed by wall relaxations. Under the $\pi$-PM field, the SNS is composed of multiple resonances centered at the odd harmonics of the average Larmor frequency $\nu_p/2$ adjusted in our experiment to be equal to ${\nu}_{L}$ of the DC field. The hight and width of the Ramsey peaks are inaccurate in the main plot due to its low frequency-resolution. From the high frequency-resolution inset, it is clear that the Ramsey peak of the $\pi$-PM field is much narrower than that of the DC field. We will only focus on the line shape of the Ramsey peaks because they correspond to the single exponential relaxation assumed in our theory and contain the information of SECs.

\begin{figure}
  \includegraphics[width=8.6cm]{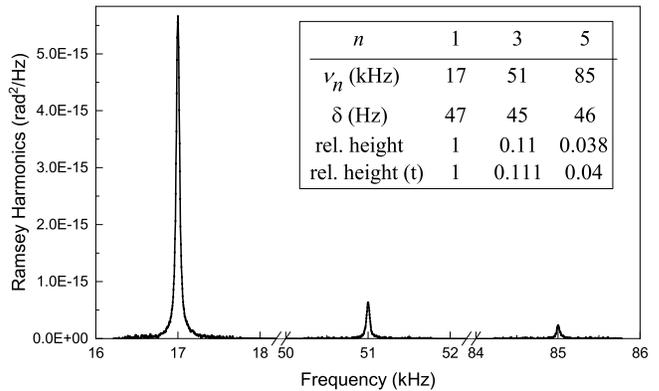}
  \caption{\label{fig:RamseyHarmonics}(color online) Joint plot of the SNS of all Ramsey harmonics in a $\pi$-PM field. Each peak is obtained by subtracting the transit pedestal from the spectrum scanned with a \SI{1.6}{kHz} frequency span. The inset table lists all the parameters for each harmonic with the last row being the theoretical relative peak heights, which are $1, 1/3^2, 1/5^2$. }
\end{figure}
Fig.\ref{fig:RamseyHarmonics} plots all the Ramsey harmonics in the $\pi$-PM field within the \SI{100}{kHz} bandwidth of the spectrum analyzer. The parameters of each peak are listed in the inset table. The measured values agree very well with the theoretical results given by Eq.(\ref{eq:PSDPiPulse}). Because $\sum_{n=1,3,5}^{\infty} 1/n^2 =\pi^2/8$, the 1st harmonic contains $\sim 81\%$ of total noise power. Thus it alone is strong enough for studying spin dynamics, while all the other harmonics are just its small replicas.

\begin{figure}
  \includegraphics[width=8.6cm]{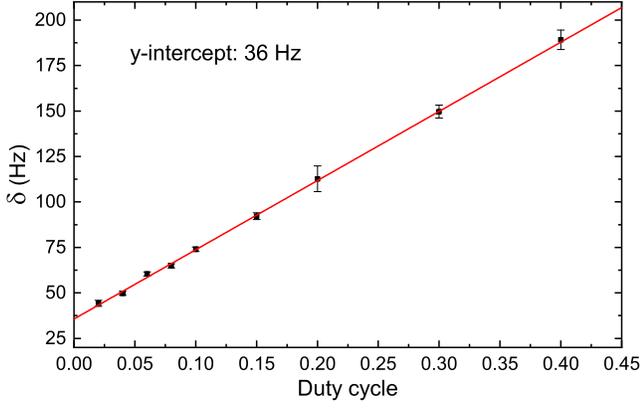}
  \caption{\label{fig:FWHMvsDutyCycle}(color online) Duty cycle dependence of the $\delta$ of the 1st Ramsey peak. The extrapolated $\delta$ at $0$ duty cycle is \SI{36}{Hz}, which should be free of any SER and solely contributed by the wall relaxation.}
\end{figure}

\begin{figure}
  \includegraphics[width=8.6cm]{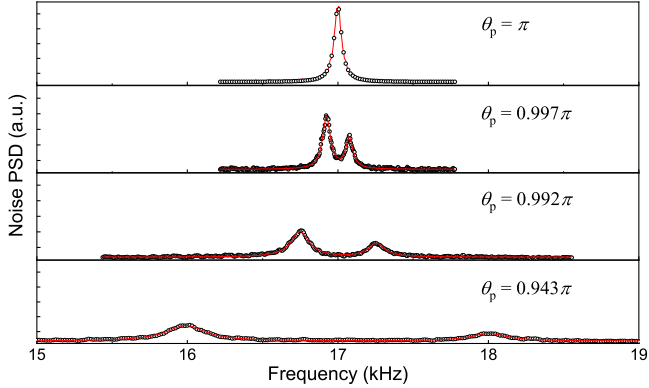}
  \caption{\label{fig:PSDvsPulseArea}(color online) SNSs with different pulse areas. The value $\theta_{p}$ is calculated from the frequency splitting of the doublet, $\nu_{n,+}-\nu_{n,-}=\nu_{p}(\theta_{p}-\pi)/\pi$. The experimental data (black circles) are fitted by a double Lorentzian of equal linewidth (red solid line). The relative height of the two peaks for each $\theta_{p}$ does not agree with that of Eq.\ref{eq:PSDArbitraryPulse} very well, especially for $\theta_{p}>\pi$. It might be caused by the asymmetric (shark-fin) shape of the actual field pulses, but the exactly reason is not understood yet.}
\end{figure}

\begin{figure}
  \includegraphics[width=8.6cm]{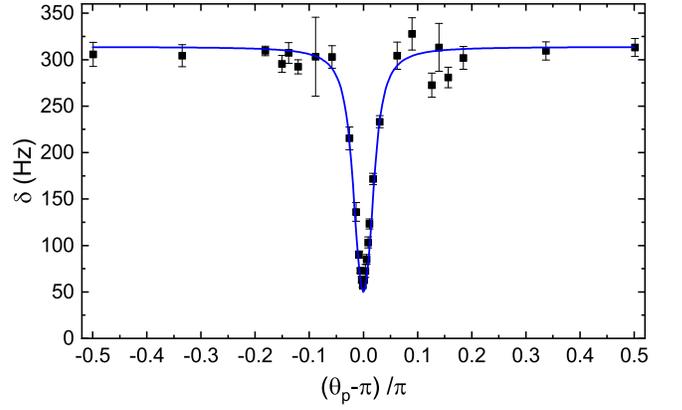}
  \caption{\label{fig:FWHMvsPulseArea}(color online) Pulse area dependence of the $\delta$ of the 1st Ramsey peak  with $d=5\%$. The Black squares are experimental values and the blue solid line is  the theoretical calculation.  The minimum $\delta$ is about \SI{50}{Hz}, contributed by the \SI{36}{Hz} wall relaxation and the \SI{14}{Hz} residue SER due to the finite pulse width.}\label{FWHMvsPluseArea}
\end{figure}

The equivalence of a PM field to a zero-field depends on two critical parameters, duty cycle $d$ and pulse area $\theta_p$.
When $\theta_{p}=\pi$, the PM field becomes more and more equivalent to a zero-field as $d\rightarrow 0$, since the probability of SECs taking place during the pulse is equal to the duty cycle. As a result, the FWHM of the Ramsey peak decreases linearly with the $d$ as is shown in Fig.\ref{fig:FWHMvsDutyCycle}.
On the other hand, as is shown in Fig.\ref{fig:PSDvsPulseArea}, for $d\ll 1$, spin dynamics and the line shape of the Ramsey peaks are governed by $\theta_p$, which can be changed in our experiment by varying the voltage of the pulse driver.
The dependence of $\delta$ on $\theta_p$ at fixed $\nu_{p}$ and $d$ is plotted in Fig.\ref{fig:FWHMvsPulseArea}. As shown in Fig.\ref{fig:ArbitraryPulseArea}(b), when $\theta_{p}$ is close but not equal to $\pi$, each pulse changes the relative angle between the atomic spins on the two hyperfine sublets by $2(\theta_{p} - \pi)$ for every time interval $T_{p}$, equivalent to a DC field with Larmor frequency, $\omega_\textrm{ef} =|\theta_{p} - \pi|/T_{p}$ as far as the evolution of this relative angle is concerned. When $\omega_\textrm{ef}$ is nonzero, the orientation of spins gets more and more diffused after each pulse when the atoms jumps back and forth between the two hyperfine states of different g-factors, causing SER broadening which saturates until $\omega_\textrm{ef}$ is much larger than the SEC rate. The exact dependence of SER on the strength of a DC field has been solved for polarized spins \cite{happer_effect_1977}. A more rigourous treatment may need the stochastic fluctuation theory outlined in Ref.\cite{roy_cross-correlation_2015}. The solid line in Fig.\ref{fig:FWHMvsPulseArea} is the calculated total broadening using \cite{happer_effect_1977} with the SER rate as the only adjustable fitting parameter. The fitted SER rate is found to be corresponding to the Rb vapor density at \SI{104}{\celsius}, which is in good agreement with our measured cell stem temperature of \SI{105.4}{\celsius}, since alkali vapor densities are usually lower in coated cells.

\begin{figure}
\includegraphics[width=8.6cm]{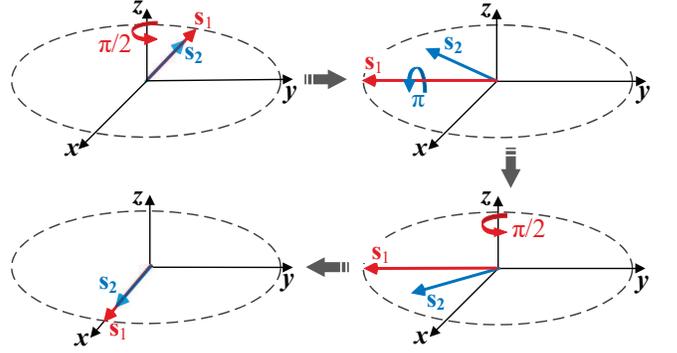}
  \caption{\label{fig:ArbitrarygFactor}(color online) A $\pi/2$ pulse for $\mathbf{S}_1$ about z-axis followed by a $\pi$ pulse for $\mathbf{S}_2$ about the y-axis followed by a $\pi/2$ pulse for $\mathbf{S}_1$ about z-axis effectively rotate both spins about the z-axis by $\pi$ regardless of the ratio of their g-factors. The probe beam is along the x direction.}
\end{figure}

This method can also be extended to study the spin correlations between spin species of arbitrary g-factors. For two spin species whose ratio of g-factors can be reduced to a ratio of two odd numbers, e.g. $n/m$, an $n\pi$-pulse for one spin is simultaneously an $m\pi$-pulse for the other, hence an effective $\pi$-pulse for both spins. Even for two spins of an arbitrary ratio of g-factors, an effective $\pi$-pulse can be formed by a sequence of pulses as illustrated in Fig.\ref{fig:ArbitrarygFactor}. Assume both spins point in the $x$-direction initially, the first pulse rotates $\textbf{S}_1$ about the z-axis by $\pi/2$ and points it to the y-axis; the second pulse rotates $\textbf{S}_2$  about the y-axis by $\pi$, causing it to lead/lag $\textbf{S}_1$ on the x-y plane by the same angle it lags/leads $\textbf{S}_1$ after the first pulse; the last pulse rotates $\textbf{S}_1$  about the z-axis by $\pi/2$ again, brings both spins to the $-x$-direction at the end. The net result is that both spins rotated a $\pi$ radian regardless of the ratio of their g-factors.

A $\pi$-PM field with duty cycle $d$ and average Larmor frequency $\bar{\nu}_L$ requires the pulse height and width to be $2\pi\bar{\nu}_L/d$ and $d/2\bar{\nu}_L$ respectively, which are limited by the coil inductance in practice, since the coil has to be large enough to guarantee the field uniformity within the sample volume and its inductance is proportional to its size. For our \SI{2}{cm} cubic cell, the preliminary current supply can create field pulses up to \SI{40}{KHz} repetition rate at $d=2.5\%$. Such a limit can be easily increased by 3 to 4 orders of magnitude for micrometer-scale samples.

In summary, we show that spin dynamics of a strictly unpolarized system in zero-field can be revealed, free from the $1/f$ noise, with SNS in a $\pi$-pulse-modulated magnetic field. In fact, any magnetic field pulse which ends up generating an overall rotation of the system while maintaining the relative orientations of the internal spins is capable of creating a macroscopically spin oscillation signal while keeping the system invariant for microscopic isotropic spin interactions. Therefore, this method is not only limited to spin-exchange-collisions, and also the $\pi$-pulse presented here is just a special case which gives the strongest macroscopic signal. The same technique also can be applied to polarized systems to suppress SER as well as any spin relaxations whose longitudinal relaxation time is longer than the transverse one, which will benefit precision measurements such as magnetometry and gyroscope.

\begin{acknowledgments}
We thanks Kangjia Liao for helpful discussions. This work is supported by National Key Research Program of China under Grant No. 2016YFA0302000, NNSFC under Grant No. 91636102, and SNSF under Grant No. 16ZR1402700. Zhang GY acknowledges the support from NNSFC under Grant No. 11704335.
\end{acknowledgments}

\bibliography{reference}

\end{document}